\documentstyle[12pt]{article}

\newcommand{\beq}{\begin{equation}}
\newcommand{\eeq}{\end{equation}}

\begin{document}

\title{Enlargement of Calderbank Shor Steane quantum codes}
\author{Andrew M. Steane\thanks
{Department of Physics, University of Oxford,
Clarendon Laboratory, Parks Road, Oxford OX1 3PU, England.}
}

\date{March 1998}
\maketitle

\begin{abstract}
It is shown that a classical error correcting
code $C = [n,k,d]$ which
contains its dual, $C^{\perp} \subseteq C$, and which can be enlarged
to $C' = [n,k' > k+1, d']$, can be converted into a quantum code of
parameters $[[n,k+k' - n, {\rm min}(d, \lceil 3d'/2 \rceil )]]$. This is a
generalisation of a previous construction, it
enables many new codes of good efficiency to be discovered. 
Examples based on classical Bose Chaudhuri Hocquenghem (BCH) codes
are discussed.
  \end{abstract} 

{\bf keywords} Quantum error correction, BCH code, CSS code
\newpage

Quantum information theory is rapidly becoming a well-established discipline.
It shares many of the concepts of classical information theory
but involves new subtleties arising from the nature of quantum mechanics
\cite{RPP}. Among the central concepts in common between classical
and quantum information is that of error correction, and the
error correcting code. Quantum error correcting codes have progressed
from their initial discovery \cite{Shor1,Steane1} and the first
general descriptions \cite{Cald1,Steane1,Steane2} to broader analyses
of the physical principles \cite{Knill,Benn,Cald2,Gott} and various code
constructions \cite{Cald2,Gott,GottP,Laf,SQC,SRM,Rains97a,Rains97b}.
A thorough discussion of the principles of quantum coding theory is offered
in \cite{CaldGF4}, and many example codes are given, together with
a tabulation of codes and bounds on the minimum distance
for codeword length $n$ up to $n=30$ quantum bits. 

For larger $n$ there is less progress, and only a few general code
constructions are known. The first important quantum code construction is
that of \cite{Cald1,Steane1,Steane2}, the resulting codes are commonly
referred to as Calderbank Shor Steane (CSS) codes. It can be shown
that efficient CSS codes exist as $n \rightarrow \infty$, but on the other
hand these codes are not the most efficient possible. I will present
here a method which permits most CSS codes to be enlarged, without
an attendant reduction in the minimum distance of the code. The resulting
codes are therefore more efficient than CSS codes. The examples
I will give are found to be among the most efficient quantum codes known,
and enable some of the bounds in \cite{CaldGF4} to be tightened.
The code construction is essentially the same as that described for
Reed-Muller codes in \cite{SRM}, the new feature is to understand how
the method works and thus prove that it remains successful for a much wider
class of code. After this some relevent theory of
Bose Chaudhuri Hocquenghem (BCH) codes \cite{BC,H,MacW} will be given and
used to construct a table of example quantum codes built by the new method.
The codes are {\em additive} and {\em pure}
in the nomenclature of \cite{CaldGF4}. A pure additive code is
{\em nondegenerate} in the nomenclature of \cite{Gott}.

\section{Quantum coding}

Following \cite{CaldGF4}, the notation $[[n,k,d]]$ is used to refer
to a quantum error correcting code for $n$ qubits
having $2^k$ codewords and minimum distance $d$.
Such a code enables the quantum information
to be restored after any set of up to $\lfloor (d-1)/2 \rfloor$ qubits
has undergone errors. In addition, when $d$ is even, $d/2$ errors can be
detected. We restrict attention to the `worst case' that any
defecting qubit (ie any qubit undergoing an unknown interaction)
might change state in a completely unknown way, so all the error
processes $X$, $Z$ and $Y=XZ$ must be correctable
\cite{Steane2,Ekert96,Knill,Gott}.

A quantum error correcting code
is an eigenspace of a commutative subgroup of the group $E$ of
tensor products of Pauli matrices. The commutativity condition can
be expressed \cite{Cald2,Gott,CaldGF4,SRM}
  \beq
H_x \cdot H_z^T + H_z \cdot H_x^T = {\bf 0}.  \label{comm}
  \eeq
where $H_x$ and $H_z$ are $(n-k \times n)$ binary matrices which
together form the {\em stabilizer} ${\cal H} = 
\left( \left. H_x \right| H_z \right)$. All vectors $(u_x | u_z)$
in the code (where $u_x$ and $u_z$ are $n$-bit strings)
satisfy $H_x \cdot u_z + H_z \cdot u_x = 0$. These
are generated by the generator  
${\cal G} = \left( \left. G_x \right| G_z \right)$ which therefore
must satisfy
   \beq
H_x \cdot G_z^T + H_z \cdot G_x^T = {\bf 0}.   \label{HG}
   \eeq
In other words ${\cal H}$ may be obtained from
${\cal G}$ by swapping the $X$ and $Z$ parts, and extracting the dual
of the resulting $(n+k) \times 2n$ binary matrix. 
The rows of $G_x$ and $G_z$ have length $n$, and
the number of rows is $n+k$. 

The weight of a vector $( u_x | u_z )$ is the Hamming weight of 
the bitwise {\sc or} of $u_x$ with $u_z$.
The minimum distance $d$ of the code $\cal C$ is the largest weight
such that there are no vectors of weight $<d$ in ${\cal C} \setminus
{\cal C^{\perp}}$, where the dual is with respect to the inner
product $((u_x | u_z), (v_x | v_z)) \equiv u_x \cdot v_z + u_z \cdot v_x$.
A {\em pure} code has furthermore no vectors of weight $<d$ in
${\cal C}$, apart from the zero vector.

The CSS code construction \cite{Cald1,Steane2} is to take
classical codes $C_1$ and $C_2$ with $C_1^{\perp} \subseteq C_2$,
and form
  \beq
{\cal G} = \left( \begin{array}{c|c} G_1 & 0 \\ 0 & G_2 \end{array}
\right), \;\;\;\;\;
{\cal H} = \left( \begin{array}{c|c} H_2 & 0 \\ 0 & H_1 \end{array}
\right).
    \label{G1G2}
  \eeq
where $G_i$ and $H_i$ are the classical generator and check matrices.
The dual condition $C_1^{\perp} \subseteq C_2$ ensures that
$H_1 \cdot H_2^T = H_2 \cdot H_1^T = 0$ and therefore the 
commutativity condition (\ref{comm}) is satisfied. If $C_1 =
[n,k_1,d_1]$ and $C_2 = [n, k_2,d_2]$ then
the minimum distance of the quantum code is ${\rm min}(d_1,d_2)$
and the number of rows in ${\cal G}$ is $k_1 + k_2$, leading
to quantum code parameters $[[n, k_1 + k_2 - n, {\rm min}(d_1, d_2)]]$.

An interesting subset of CSS codes is that given by the above
construction starting from a classical $[n,k,d]$ which contains
its dual, leading to a quantum $[[n,2k-n,d]]$ code.

\section{New code construction}

I will present the new construction by stating and proving the following.

{\bf Theorem 1}. {\em Given a classical binary error correcting
code $C = [n,k,d]$ which
contains its dual, $C^{\perp} \subseteq C$, and which can be enlarged
to $C' = [n,k' > k+1, d']$, a pure quantum code of
parameters $[[n,k+k' - n, {\rm min}(d, \lceil 3d'/2 \rceil )]]$ can be constructed.}

{\bf Proof}. The generator for the quantum code is
  \beq
{\cal G} = \left( \begin{array}{c|c} 
D   & A D \\
G   & 0 \\
  0 & G  \end{array}
\right),
  \eeq
where $G$ generates the classical code $C$, and
$G$ and $D$ together generate $C'$, as does $G$ and $A D$ together
(we will choose $A$ such that $D$ and $A D$ generate the same set).

The stabilizer is
  \beq
{\cal H} =  \left( \begin{array}{c|c} 
\tilde{A} B   & B \\
H'   & 0 \\
0 & H'  \end{array}
\right),
  \eeq
where $H'$ checks the code $C'$, so has $n-k'$ rows, $\{H', B \}$
checks the code $C$, so $B$ has $k'-k$ rows, and
  \beq
\tilde{A} = B D^T \left( A^T \right)^{-1} \left( B D^T \right)^{-1}
  \eeq

From the dual conditions specified in the theorem,
$H' H'^T = 0$ and $H' B^T = 0$ so the commutativity condition
(\ref{comm}) is satisified. The definition of $\tilde{A}$ ensures we have
the correct stabilizer since
  \beq
\tilde{A} B (A D)^T = B D^T.
  \eeq

Since the number of rows in the generator is $k + k'$, the dimension
of the quantum code is $k + k' - n$. It remains to prove that the
minimum distance is ${\rm min}(d, \lceil 3d'/2 \rceil )$. 

We choose $A$ such that $D$ and $A D$ generate the same set. Therefore
for any vector
$(u|v)$ generated by $(D | A D)$, either $u = v$
or ${\rm wt}(u + v) \ge d'$. We choose the map $A$ such that
$u = v$ never occurs (a fixed point free map). This can be
achieved as long as $D$ has more than one row, by, for example, the map
  \beq
A = \left( \begin{array}{c}
0100 \ldots 0 \\
0010 \ldots 0 \\
0001 \ldots 0 \\
\ldots        \\
0000 \ldots 1 \\
1100 \ldots 0
\end{array} \right).
  \eeq

To complete the proof we will show that for any non-zero vector
$(u | v)$ generated by ${\cal G}$, ${\rm wt}(u | v) \ge {\rm min}(d,3d'/2)$
(and therefore ${\rm wt}(u | v) \ge {\rm min}(d,\lceil 3d'/2 \rceil )$.)

For the non-zero vector $(u | v)$, if either 
${\rm wt}(u) \ge d$ or ${\rm wt}(v) \ge d$
then ${\rm wt}(u | v) \ge d$, so the conditions of the
theorem are satisfied. The only remaining vectors are those for which
both ${\rm wt}(u) < d$ and ${\rm wt}(v) < d$.
Now, ${\rm wt}(u)$ can
only be less than $d$ if $D$ is involved in the generation of
$u$, and ${\rm wt}(v)$ can
only be less than $d$ if $A D$ is involved in the generation of
$v$, since $G$  on its own generates a binary code of minimum
distance $d$. However, since the map $A$ is fixed-point free,
and using the fact that $D$ and $AD$ generate the same set, 
the binary vector $u+v$ is not zero and is a member of a distance $d'$
code, therefore ${\rm wt}(u+v) \ge d'$. We thus have the
conditions $\{ {\rm wt}(u) \ge d'$, ${\rm wt}(v) \ge d'$, 
${\rm wt}(u+v) \ge d' \}$. These are sufficient to imply that
${\rm wt}(u|v) \ge 3 d'/2$. For, if $u$ and $v$ overlap in $p$
places, then ${\rm wt}(u+v) = {\rm wt}(u) - p + {\rm wt}(v) - p$
and ${\rm wt}(u|v) = {\rm wt}(u) + {\rm wt}(v) - p$
$=({\rm wt}(u) + {\rm wt}(v) + {\rm wt}(u+v))/2$ $\ge 3d'/2$. 
This completes the proof.

The above construction was applied to Reed-Muller codes in \cite{SRM}.
These codes are not very efficient (they have small $k/n$ for given
$n,d$) but they have the advantage of being easily decoded. A large
group of classical codes which combine good efficiency with ease
of decoding are the BCH codes. They include Reed Solomon codes as
a subset. I will now derive a set of quantum error
correcting codes from binary BCH codes using the above construction,
combined with some simple BCH coding theory.

\section{Application to binary BCH codes}

Properties of BCH codes are discussed and proved in, for example,
\cite{MacW}. A binary BCH code of designed distance $\delta$
is a cyclic code of length $n$ over GF(2) with generator polynomial
  \beq
g(x) = {\rm l. c. m.} \{ M^{(b)}(x), M^{(b+1)}(x), \ldots
, M^{(b+\delta-2)}(x) \}  \label{gx}
  \eeq
where
  \beq
M^{(s)}(x) = \prod_{i \in C_s} \left(x-\alpha^i \right),  \label{Ms}
  \eeq
in which $\alpha$ is a primitive $n$th root of unity over GF(2),
and $C_s$ is a cyclotomic coset mod $n$ over GF(2), defined by
  \beq
C_s = \{ s, 2s, 4s, \ldots, 2^{m_s - 1} s \},  \label{Cs}
  \eeq
where $m_s = |C_s|$ is obtained from $2^{m_s} s \equiv s \bmod n$.
The dimension of the code is $k = n - \deg (g(x))$. From (\ref{gx})
and (\ref{Ms}) this implies
$k = n - \sum_s |C_s|$ where the sum ranges from $s=b$ to $s=\delta+b-2$
but only includes each distinct cyclotomic coset once. This can also
be expressed $k = n - |{\cal I}_C|$ where 
${\cal I}_C = C_b \cup C_{b+1} \cup \cdots \cup C_{b+\delta-2}$
is called the {\em defining set}. The minimum
distance of the code is $d \ge \delta$. 

The dual of a cyclic code is cyclic. Grassl {\em et al.} \cite{Grassl}
derive the useful criterion that a cyclic code contains its dual
if the union of cyclotomic cosets contributing to $g(x)$ does
not contain both $C_s$ and $C_{n-s}$. In other words
  \beq
\{ (n-i) \not\in {\cal I}_C \; \; \forall i \in {\cal I}_C \}
\Rightarrow C^{\perp} \subseteq C.     \label{cond}
  \eeq

\subsection{Primitive BCH codes}

Consider first the BCH codes with $n = 2^m - 1$, the so-called
primitive BCH codes. In order to find the codes which satisfy
the condition (\ref{cond}), we will find the smallest $s$ such
that $n-r \in C_s$ for some $r \le s$. The largest permissible designed
distance will then be $\delta = s$. For even $m$,
the choice $s=2^{m/2} - 1$ gives
$s 2^{m/2} = n - s \Rightarrow C_s = C_{-s}$,
so this is an upper bound on $s$. For odd
$m$, an upper bound is provided by $s = 2^{(m+1)/2} - 1$ since
then $s 2^{(m-1)/2} = n - (s-1)/2$. We will
show that these upper bounds can be filled, i.e. that no smaller
$s$ leads to $n-r \in C_s$ for $r \le s$. 

For $n=2^m - 1$, the elements of the cyclotomic cosets $C_s$ are largest when
$s$ is one less than a power of 2, 
$s = 2^j - 1$. Specifically, for $s = 2^j - 1$ we have
$(s 2^i \bmod n) > (r 2^i \bmod n)
\; \forall i<m, r<s$. This is obvious for $s 2^i < n$ and
the proof for $s 2^i > n$ is straightforward. The largest element in
$C_s$ ($s=2^j - 1$) is obtained for the largest $i$
such that $s 2^i < n$, giving $\max(C_s) = 2^m - 2^{m-j} =
n - r$ where $r = 2^{m-j} - 1$. This element
$\max(C_s) = n-r$ is the largest in the defining set ${\cal I}_C$
for a code of designed distance $\delta = s$,
therefore it is only possible for ${\cal I}_C$ to contain both $i$ and $n-i$
(for any $i$) if it contains $r$ and $n-r$, since
$r = 2^{m-j} - 1$ is the smallest element in its coset, and
any other pairs $i, n-i$ must have $i > r$.
Finally, we have a failure of the condition (\ref{cond}) only
if $r \le s$, that is $2^{m-j} - 1
\le 2^j - 1$, therefore $j \ge \lceil m/2 \rceil$. 

To summarise the above, we have the proved following:

{\bf Lemma}: {\em The primitive binary BCH codes contain their duals
if and only if the designed distance satisfies}
  \beq
\delta \le 2^{\lceil m/2 \rceil} -1
  \eeq

Using the code construction of theorem 1, together with this lemma,
the list of quantum codes in table 1 is obtained. The further property
used is that BCH codes are nested, i.e. codes of smaller distance
contain those of larger, which is obvious since the former can be obtained
from the latter by deleting parity checks. The first entry for each value
of $n$ uses $\{C =$ extended BCH code$\}$ with $\{C'=$ even weight code$\}$
to obtain a distance 3 quantum code. The codes of larger distance involve
only BCH codes, for these a quantum code is obtained both from the unextended
and extended versions. The parameters $[[n,K,D]]$ given in the table
are for the extended BCH codes (i.e. extended by an overall parity check).
Using unextended codes leads to a further quantum code of parameters
$[[n-1,K+1,D-1]]$, for $D>3$. 

\subsection{Non-primitive BCH codes}

When $n \ne 2^m - 1$ the cyclotomic cosets mod $n$ do not have so much
structure so in general the only way to find if condition (\ref{cond})
is satisfied is to examine each coset individually.

One way in which the requirement (\ref{cond}) is not met is if
$C_s$ contains both $i$ and $-i \bmod n$, which implies $C_s = C_{-s}$,
for some $C_s \subseteq {\cal I}_C$.
If $s$ is the smallest element in $C_s$, then $i,n-i \in C_s$ if and only
$s,n-s \in C_s$, from which $s 2^j \equiv -s \bmod n$ for some $j < m_s$.
Multiplying by $2^j$ we have $s 2^{2j} \equiv -s 2^j \equiv s \bmod n$,
therefore $j = m_s/2$ and this is only possible for even $m_s$.
Furthermore, since $m_{s \ge 1}$ is a factor of $m_1$, $m_s$
can be even only if $m_1$ is even. This observation slightly reduces
the amount of checking to be done.

The values of $n$ in the range $1 < n \le 127$
for which $C_1$ does not contain $n-1$ are $\{$ 7,15,21,23,31,35,39,45,
47,49,51,55,63,69,71,73,75,77,79,85,87,89,91,93,95,103,
105,111,115,117,119,121,123,127$\}$.
An efficient code is obtained
if one or more of the cosets is small,
this happens for $n =$ 21, 23, 45, 51, 73, 85, 89,
93, 105, 117 (not counting primitive codes). Quantum codes obtained
from BCH codes with these values of $n$ are listed in table 2. Further
good codes exist in the range $127 < n < 511$
for $n =$ 133, 151, 153, 155, 165, 189, 195, 217, 219,
255, 267, 273, 275, 279, 315, 337, 341, 381, 399, 455.

\section{Efficiency}

The code parameters in tables 1 and 2 compare well with the most
efficient quantum codes known. For example, the $[[22,5,6]]$
code permits some of the lower existence bounds in \cite{CaldGF4}
to be raised, and the $[[32,15,6]]$  and $[[32,5,8]]$ codes fill
lower existence bounds. The $[[93,68,5]]$ code is comparable
with the $[[85,61,5]]$ code quoted in \cite{CaldGF4}, though
the $[[93,53,7]]$ code is not as good as $[[85,53,7]]$ quoted
in \cite{CaldGF4}. Obviously the quantum codes based on BCH codes will
be best for primitive BCH codes, so we expect the codes in table 1
rather than table 2 to compare best with other code constructions.
Indeed, the distance 3 codes in table 1 are the previously known
Hamming codes \cite{Gott,SRM,CaldGF4} and are optimal.

The quantum codes constructed by theorem 1 have an upper bound on
the rate $K/n = (k + k')/n - 1$ arising from the upper bound
on $k$ and $k'$ for binary codes. In the asymptotic limit this
bound on the quantum codes is 
  \beq
K / n < R(d/n) + R(2d/3n) -1,   \label{bound}
  \eeq
where $R(d/n)$ is the maximum rate of a binary $[n,k,d]$ linear code.
For example the sphere-packing bound is $R(d/n) < 1 - H(d/2n)$;
the codes we have discussed have parameters lying close to
this bound (though in the limit of large $n$ it is known that BCH
codes are no longer efficient). Taking $R(x)$ equal to the 
McEliece-Rodemich-Rumsey-Welch upper bound \cite{McE77}, we find
$K/n \ge 0$ for $d/n < 0.2197$ in the limit of large $n$.
This may be compared with
$d/n < 0.1825$ for CSS codes and the limit $d/n < 0.308$
for pure quantum stabilizer codes discussed by 
Ashikhmin \cite{Ashik}.

The author is supported by the Royal Society and by St Edmund Hall, Oxford.

\newpage

\begin{tabular}{r|rrrr|rr}
n & k & k' & d & d' & K & D\\
\hline
 8 &  4 &  7 & 4 & 2 & 3 & 3 \\
\\
16 & 11 & 15 & 4 & 2 & 10 & 3 \\
\\
32 & 26 & 31 & 4 & 2 & 25 & 3 \\
32 & 21 & 26 & 6 & 4 & 15 & 6 \\
32 & 16 & 21 & 8 & 6 &  5 & 8 \\
\\
64 & 57 & 63 & 4 & 2 & 56 & 3 \\
64 & 51 & 57 & 6 & 4 & 44 & 6 \\
64 & 45 & 51 & 8 & 6 & 32 & 8 \\
\\
128 & 120 & 127 & 4 & 2 & 119 & 3 \\
128 & 113 & 120 & 6 & 4 & 105 & 6 \\
128 & 106 & 113 & 8 & 6 & 91 & 8 \\
128 & 99 & 113 & 10 & 6 & 84 & 9 \\
128 & 92 & 106 & 12 & 8 & 70 & 12 \\
128 & 85 & 99 & 14 & 10 & 56 & 14 \\
128 & 78 & 99 & 16 & 10 & 49 & 15 \\
\\
256 & 247 & 255 & 4 & 2 & 246 & 3 \\
256 & 239 & 247 & 6 & 4 & 230 & 6 \\
256 & 231 & 239 & 8 & 6 & 214 & 8 \\
256 & 223 & 239 & 10 & 6 & 206 & 9 \\
256 & 215 & 231 & 12 & 8 & 190 & 12 \\
256 & 207 & 223 & 14 & 10 & 174 & 14 \\
256 & 199 & 223 & 16 & 10 & 166 & 15
\end{tabular}

Table 1. Parameters $[[n,K,D]]$ of the quantum codes obtained from
primitive binary BCH codes, for $n \le 256$. The BCH codes have been
extended by an overall parity check in order to allow the distance
3 quantum code to be obtained by combining a BCH code
with the even weight code. For $D>3$ if the unextended BCH codes
are used, a $[[n-1,K+1,D-1]]$ quantum code is obtained.

\newpage

\begin{tabular}{r|rrrr|rr}
n & k & k' & d & d' & K & D\\
\hline
22 & 15 & 21 & 4 & 2 & 14 & 3 \\
22 & 12 & 15 & 6 & 4 & 5 & 6 \\
\\
46 & 33 & 45 & 4 & 2 & 32 & 3 \\
46 & 29 & 33 & 6 & 4 & 16 & 6 \\
\\
52 & 43 & 51 & 4 & 2 & 42 & 3 \\
\\
74 & 64 & 73 & 4 & 2 & 63 & 3 \\
74 & 55 & 64 & 6 & 4 & 45 & 4 \\
74 & 46 & 55 & 10 & 6 & 27 & 9 \\
\\
86 & 77 & 85 & 4 & 2 & 76 & 3 \\
86 & 69 & 77 & 6 & 4 & 60 & 6 \\
\\
90 & 78 & 89 & 4 & 2 & 77 & 3 \\
90 & 67 & 78 & 6 & 4 & 55 & 6 \\
90 & 56 & 67 & 10 & 6 & 33 & 9 \\
90 & 45 & 56 & 12 & 10 & 11 & 12 \\
\\
94 & 83 & 93 & 4 & 2 & 82 & 3 \\
94 & 78 & 83 & 6 & 4 & 67 & 6 \\
94 & 68 & 78 & 8 & 6 & 52 & 8 \\
94 & 58 & 78 & 10 & 6 & 42 & 9 \\
94 & 53 & 68 & 12 & 8 & 27 & 12 \\
\\
106 & 93 & 104 & 4 & 2 & 92 & 3 \\
106 & 81 & 93 & 6 & 4 & 68 & 6 \\
106 & 75 & 81 & 8 & 6 & 50 & 8 \\
106 & 71 & 81 & 10 & 6 & 46 & 9 \\
\\
118 & 105 & 117 & 4 & 2 & 104 & 3 \\
118 & 93 & 105 & 6 & 4 & 80 & 6 \\
118 & 81 & 93 & 8 & 6 & 56 & 8 \\
118 & 69 & 93 & 10 & 6 & 44 & 9
\end{tabular}

Table 2. As table 1, but for non-primitive BCH codes with $n < 127$.
 

\begin{thebibliography}{12}

\bibitem{RPP} For a general introduction to quantum information theory, see,
for example, A. M. Steane, ``Quantum computing'',
{\em Rep. Prog. Phys.}, to be published (preprint quant-ph/9708022).

\bibitem{Shor1} P. W. Shor, 
``Scheme for reducing decoherence in quantum computer memory,''
{\em Phys. Rev. A},  vol. 52, pp. R2493-R2496, Oct. 1995.

\bibitem{Steane1} A. M. Steane, 
``Error correcting codes in quantum theory,''
{\em Phys. Rev. Lett.}, vol. 77, pp. 793-767, July 1996.

\bibitem{Cald1} A. R. Calderbank and P. W. Shor, 
``Good quantum error-correcting codes exist,''
{\em Phys. Rev. A}, vol. 54, pp. 1098-1105, Aug. 1996.

\bibitem{Steane2} A. M. Steane, 
``Multiple particle interference and quantum error correction,''
{\em Proc. Roy. Soc. Lond. A}, vol. 452, pp. 2551-2577, Nov. 1996.

\bibitem{CaldGF4} A. R. Calderbank, E. M. Rains, P. W. Shor
and N. J. A. Sloane, 
``Quantum error correction via codes over $GF(4)$'',
{\em IEEE Trans. Information Theory}, to be published.
(LANL eprint quant-ph/9608006).

\bibitem{Ekert96} A. Ekert and C. Macchiavello,
``Quantum error correction for communication,''
{\em Phys. Rev. Lett.}, vol. 77, pp. 2585-2588, Sept. 1996.

\bibitem{Knill} E. Knill and R. Laflamme,
``A theory of quantum error correcting codes,''
{\em Phys. Rev. A}, vol. 55, pp. 900-911, (1997). 

\bibitem{Laf} R. Laflamme, C. Miquel, J. P. Paz and W. H. Zurek,
``Perfect quantum error correcting code,''
{\em Phys. Rev. Lett.}, vol. 77, pp. 198-201, July 1996.

\bibitem{Benn} C. H. Bennett, D. P. DiVincenzo, J. A. Smolin and
W. K. Wootters,
``Mixed state entanglement and quantum error correction,''
{\em Phys. Rev. A}, vol. 54, pp. 3822-3851, (1996).

\bibitem{Cald2} A. R. Calderbank, E. M. Rains, N. J. A. Sloane
and P. W. Shor, 
``Quantum error correction and orthogonal geometry,''
{\em Phys. Rev. Lett.}, vol. 78, pp. 405-409, (1997).

\bibitem{SQC} A. M. Steane, 
``Simple quantum error correcting codes,''
{\em Phys. Rev. A}, vol. 54, pp. 4741-4751, (1996). 

\bibitem{Gott} D. Gottesman, 
``Class of quantum error-correcting codes saturating the quantum
Hamming bound,''
{\em Phys. Rev. A}, vol. 54, pp. 1862-1868, (1996).

\bibitem{Rains97a} E. M. Rains,
``Nonbinary quantum codes,''
Preprint quant-ph/9703048.

\bibitem{Rains97b} E. M. Rains,
``Quantum codes of minimum distance two,''
Preprint quant-ph/9704043.

\bibitem{SRM} A. M. Steane,
``Quantum Reed-Muller codes,''
{\em IEEE Trans. Inf. Theory}, to be published
(preprint quant-ph/9608026).

\bibitem{GottP} D. Gottesman,
``Pasting quantum codes,''
Preprint quant-ph/9607027.

\bibitem{BC}
R. C. Bose and C. R. Ray-Chaudhuri,
``On a class of error-correcting binary group codes,''
{\em Information and Control}, vol. 3, pp/ 68-79.

\bibitem{H}
A. Hocquenghem, 
``Codes correcteurs d'erreurs,''
{\em Chiffres}, vol 2, pp. 147-156, September, 1959.

\bibitem{Grassl}
M. Grassl, Th. Beth and T. Pellizzari,
``Codes for the quantum erasure channel,''
{\em Phys. Rev. A}, vol 56, pp. 33-38 (1997).

\bibitem{MacW} F. J. MacWilliams and N. J. A. Sloane, 
{\em The Theory of Error-Correcting Codes,}
Amsterdam: North-Holland, 1977. 

\bibitem{McE77}
R. J. McEliece, E. R. Rodemich, H. C. Rumsey, Jr. and L. R. Welch,
``New upper bounds on the rate of a code via the Delsart-MacWilliams
inequalities,''
{\em IEEE Trans. Inf. Theory}, vol 23, pp. 157-166 (1977).

\bibitem{Ashik} A. Ashikhmin,
``Remarks on bounds for quantum codes,''
preprint quant-ph/9705037.


\end{thebibliography}
\end{document}